\documentclass[
  journal=largetwo,
  manuscript=research-paper,
  year=2020,
  volume=37,
]{cup-journal}

\usepackage{amsmath}
\usepackage{amssymb}
\usepackage{physics}
\usepackage[nopatch]{microtype}
\usepackage{booktabs}
\usepackage{aas-macros}
\usepackage{wasysym}

\title{Early Accretion Onset in Long-Period Isolated Pulsars}

\author{M.D. Afonina}

\affiliation{Sternberg Astronomical Institute, Lomonosov Moscow State University, Universitetskij pr. 13, 119234, Moscow, Russia}
\alsoaffiliation{Faculty of Physics, Lomonosov Moscow State University, 1/2 Leninskie Gory, Moscow, 119991, Russia}

\email[M.D. Afonina]{afonina.md19@physics.msu.ru}

\author{A.V. Biryukov}
\affiliation{Sternberg Astronomical Institute, Lomonosov Moscow State University, Universitetskij pr. 13, 119234, Moscow, Russia}
\alsoaffiliation{Faculty of Physics, HSE University, 21/4 Staraya Basmannaya str., Moscow, 105066, Russia}
\alsoaffiliation{Kazan Federal University, 18 Kremlyovskaya str., Kazan, 420008, Russia}

\author{S.B. Popov}
\affiliation{Sternberg Astronomical Institute, Lomonosov Moscow State University, Universitetskij pr. 13, 119234, Moscow, Russia}
\alsoaffiliation{ICTP–Abdus Salam International Center for Theoretical Physics, Strada Costiera 11, I-34151 Trieste, Italy}

 %% First letter not capped
\begin{document}

\begin{abstract}
% 400 words or less
We model long-term magneto-rotational evolution of isolated neutron stars with long initial spin periods. This analysis is motivated by the recent discovery of young long-period neutron stars observed as periodic radio sources: PSR J0901-4046, GLEAM-X J1627-52, and GPM J1839-10. Our calculations demonstrate that for realistically rapid spin-down during the propeller stage isolated neutron stars with velocities $\lesssim100$~km~s$^{-1}$ and assumed long initial spin periods can reach the stage of accretion from the interstellar medium within at most a few billion years as they are born already at the propeller stage or sufficiently close to the critical period of the ejector-propeller transition. If neutron stars with long initial spin periods form a relatively large fraction of all Galactic neutron stars then the number of isolated accretors is substantially larger than it has been predicted by previous studies. 
\end{abstract}

%\noindent sample  text 

\section{Introduction}

 Neutron stars (NSs) are a natural product of stellar evolution. The main channel of NS formation is related to core-collapse supernovae (SN). The rate of core-collapse SN in the Milky Way now is about 1/50 yrs$^{-1}$ (\cite{2021NewA...8301498R}). %2009.03438 
This number provides a rough estimate $\sim$~few$\times10^8$ NSs in the Galaxy. Today we know about several thousand NS. Mostly, known NSs are radio pulsars (\cite{2005AJ....129.1993M}). %ATNF. 
A few hundred NSs are observed in accreting binary systems (\cite{2007A&A...469..807L, 2023A&A...671A.149F, 2023arXiv230316137N, 2023A&A...675A.199A}).  %HMXB and LMXB catalogues 2303.16137 2302.02656 0707.0544 2303.16168
All other known NSs~--- magnetars, central compact objects in supernova remnants (CCOs), isolated cooling NSs, etc.~--- see a review by  \citet{2023Univ....9..273P},~--- provide less than one hundred objects altogether.  

The majority of Galactic NSs are old objects. Their radio pulsar activity (which has a typical duration $\sim10^7$~yrs, \cite{2018PhyU...61..353B}) ceased, they are too cold to be observed due to their surface thermal emission, and they do not demonstrate magnetar activity (typical ages of magnetars are $<10^5$~yrs). 
Still, there is a way to `resurrect' such a compact object~--- accretion from the interstellar medium (ISM).

 The existence of accreting isolated NSs (AINSs) was proposed more than half a century ago (\cite{1970ApL.....6..179O, 1971SvA....14..662S}). However, despite many attempts, see e.g. \citet{2010ApJ...714.1424T} and references therein,  not a single robust candidate has ever been proposed.
 High hopes were placed on ROSAT observations (see, e.g. \citet{2000PASP..112..297T} about AINS studies before the year 2000).  Negative results of AINS searches in the ROSAT data were explained by \citet{2000ApJ...530..896P}. Due to the large kick velocities that NSs obtain at birth, the onset of accretion is significantly delayed. Under standard at that moment assumptions, \citet{2000ApJ...530..896P} obtained that at most a few percent of old NSs can be observed as accretors, and expected fluxes of most of them are low even for high efficiency of accretion.

  The number of AINSs was re-calculated by \citet{2010MNRAS.407.1090B} accounting for a significant fraction of highly magnetized NSs and field decay. It was demonstrated that NSs with a high initial magnetic field reach the stage of accretion more rapidly as they spin down faster while the field is high. As only low-velocity NSs in relatively dense regions of the ISM can become accretors within the present Galactic age, in the solar vicinity the fraction of accretors is high: $\sim 35-40\%$ (or even twice higher if subsonic propellers are included); high-velocity NSs spend most of their lives high above the Galactic plane in low-density regions. 

  The physics of accretion at low rates onto magnetized NSs contains many uncertainties.  
That is why, it was many times analyzed numerically (see e.g., \cite{2012ApJ...752...30B, 2012MNRAS.420..810T} and references therein). Still, analytical approaches are very much welcomed, as they allow us to calculate easily the properties of sources with various parameters. 
  A new analytical model of spherical accretion at low rates~--- so-called settling accretion,~--- has been proposed not long ago by \citet{2012MNRAS.420..216S}. 
\citet{2015MNRAS.447.2817P} applied this model to the case of AINSs. 
These authors concluded that AINS can appear as transient quasi-periodic sources as matter can be accumulated in an envelope around the NS, but the mean luminosity is much lower than predicted by the Bondi formula.

 Many uncertainties in the physics of low-rate accretion can be removed if AINSs are finally discovered. 
 Detection of AINSs will be of great value for many reasons. E.g., this is the best way to probe the magneto-rotational evolution of NSs on the time scale of about a few Gyrs. In particular, long-term magnetic field decay can be probed in this way.

 The behavior and observational appearance of old NSs also depend on the initial parameters and early evolution. 
As we mentioned above, the role of kick velocities and initial magnetic fields was studied, already. 
Initial spin periods never were considered as crucial ingredients because it is typically assumed that they are always short enough~--- $\lesssim 1$~s (e.g., \cite{2012Ap&SS.341..457P}),~--- to provide a common start for the magneto-rotational evolution. However, the recent discovery of the 76-second radio pulsar PSR J0901-4046 (\cite{2022NatAs...6..828C}) challenges this assumption. In addition, two radio sources with periods $\sim1000$~s were found: GLEAM-X J1627-52 (\cite{2022Natur.601..526H}) and GPM J1839-10 (\cite{2023Natur.619..487H}). The nature of these objects is not certain (their properties are well summarized e.g., by \cite{2023MNRAS.520.1872B}). Still, it is quite plausible that they are young NSs, and the observed periodicity is due to their spin. 

  The origin of long periods is not known, yet. One of the realistic possibilities is related to an episode of fall-back accretion and rapid spin-down of an NS due to interaction with the surrounding matter (\cite{2022ApJ...934..184R}). Another possibility is related to a rapid spin-down of magnetars due to winds, as suggested by  \cite{2022MNRAS.517.3008P}. In this model, the spin period can increase by a factor $\sim \exp(10)$ during the early cooling stage lasting for $\sim 100$~s. % Prasanna et al. 2022 
  In the model by \citet{2022ApJ...934..184R} long spin periods are achieved on a much longer time scale $\sim 10^4$~yrs. Still, it is much shorter than the typical duration of the ejector stage. Thus, we consider spin periods after the phase of a rapid spin-down is over, as initial.
%  Then, it is fair to say that such NSs have long initial spin periods because spin-down is very rapid. 
  If all three long-period radio sources are young NSs then the fraction of such objects can be non-negligible. Then, it is important to consider how long initial spin periods can influence the fate of old isolated NSs. 

%\cite{2023MNRAS.520.1872B} % Beniamini et al. 2022

 In this paper, we discuss the long-term evolution of NSs, similar to PSR J0901-4046, which have long spin periods already in their youth. For different combinations of parameters, we calculate the age when such NSs can start to accrete from the ISM. 
 In the next section, the general properties of spin evolution of isolated neutron stars are reviewed. In Sec.~3 we describe the specific model of NS magneto-rotational evolution adopted in our work.
 Then, in Sec.~4 we present results of the calculations, which are discussed in Sec.~5.
 %In the next section, our model of magneto-rotational evolution of NSs is presented. In Sec.~3 we present the results of the calculations. Then, in Sec.~4 discussion of the model and obtained results is given. 
We conclude, summarizing our findings, in section 6. 

%2201.11704 Long-period Pulsars as Possible Outcomes of Supernova Fallback Accretion
%2212.10501 Evolution of the long-period pulsar PSR J0901-4046

%2307.10351    A long-period radio transient active for three decades: population study in the neutron star and white dwarf rotating dipole scenarios

%2210.09323     Evidence for an abundant old population of Galactic Ultra long period magnetars and implications for fast radio bursts

\section{Spin evolution of isolated neutron stars}

As an isolated neutron star (INS) evolves, it loses rotational energy and interacts with the surrounding material. This results in changing regimes of its behavior and observational appearance. In this section, we aim to give a brief overview of INS evolutionary stages~--- ejector, propeller, and accretor~--- and the transitions between them in terms of critical radii and critical periods. 
% as it is considered in~\cite{2022MNRAS.511.4447K}.

% может перечислить все параметры? Это неизменяемые M, I и диапазоны B, P, v и 
 Let consider the spin evolution of an %spherical 
INS with moment of inertia $I$, mass $M$, spin velocity $\omega$ and magnetic moment $\mu$. The star is moving with the speed $v_\infty$ through the interstellar medium of number density $n$ and sound velocity $c_{\text{s}}$.

The general equation describing the spin evolution of such a NS is the Euler equation:
\begin{equation}
\label{Euler}
    I\dv{\omega}{t} = -K,
\end{equation}
where $K$ is the spin-down torque acting on the star at a specific stage.

 %We consider the evolution of the spin period $P$ of an INS with mass $M = 1.4~M_{\astrosun}$, radius $R_{\text{NS}} = 10$~km, and moment of inertia $I = 10^{45}\text{~g~cm}^2$. The INS has a magnetic moment $\mu = B R_{\text{NS}}^3$, where $B$ is a magnetic field strength at the equator.
 %We will also use the angular frequency $\omega = 2 \pi / P$ in order to make the formulae more compact. The surrounding medium of the NS is an ISM with sound velocity $c_{\text{s}}$ and density $\rho = n m_{\text{p}}$, where $n=0.3$~cm$^{-1}$ is a number density and $m_{\text{p}}$ is a proton mass. 

 % Общее уравнение для потерь энергии в виде I \omega_dot = - K, убрать все остальные формы
 % кси оставить тут, а расшифровку убрать в модель
 % Можно сказать о свойствах межзвездной среды только перед радиусом Шварцмана, где она первый раз появляется
\subsection{Ejector}

%According to the standard scenario 
%proposed by \cite{1971SvA....15..342S} and \cite{1975A&A....39..185I} 
%an NS should initially be an ejecting pulsar (Ejector). expelled
During the ejector evolutionary stage, the ambient matter of the ISM remains far beyond the light cylinder radius $R_{\text{l}}$, which represents the maximum distance at which closed magnetic field lines can exist:

\begin{equation}
    \label{R_l}
    R_{\text{l}}=\frac{c}{\omega},
\end{equation}
where $c$ is the light speed. 

At a distance larger than $R_{\text{l}}$, the low-frequency electromagnetic radiation and the flow of relativistic particles sweep the matter out, so the magnetosphere and the surrounding matter do not interact directly. At this stage, the NS slows down due to the ejection of these magnetized winds. The corresponding spin-down torque $K_{\text{E}}$ is:

\begin{equation} 
    \label{K_E}
        K_{\text{E}}=\xi\frac{\mu^2}{R_{\text{l}}^3}.
\end{equation}
%where $\xi$ is a dimensionless coefficient of the order of unity.
From the numerical simulation of the plasma-filled magnetosphere, \citet{2006ApJ...648L..51S} obtained
\begin{equation}
    \xi = k_0 + k_1\sin^2\alpha,
\end{equation}
where $\alpha$ is the angle between the magnetic and spin axes. Later, \citet{2014MNRAS.441.1879P} estimated $k_0 \approx 1$ and $k_1 \approx 1.4$ and showed that their values are weakly dependent on the NS parameters. Note, that often the braking torque at the ejector stage is taken to be smaller. Thus, in our model an NS spins down more effectively and the ejector stage is shorter than in some previous calculations, e.g. \citet{2010MNRAS.407.1090B}.

Even though during the ejector stage, the NS magnetosphere does not interact with the surrounding material, the ISM is influenced by the NS presence. The characteristic radius of the gravitational influence of the NS is the gravitational capture radius
 %We accept $n=0.3$~cm$^{-1}$ for the mean value of the ISM number density.
%\frac{2 G M}{v_{\infty}^2 + c_{\text{s}}^2} = 
\begin{equation}
    \label{R_G}
    R_{\text{G}}=\frac{2 G M}{v^2},
\end{equation}
where $G$ is the Newton constant and $v^2 = v_\infty ^2 + c_{\text{s}}^2$. %, where $v_\infty$ is the NS velocity relative to the ISM.

The amount of interstellar matter that is captured by the NS per unit of time is $\dot{M}$. Following \citet{1944MNRAS.104..273B} we assume
%as the main characteristic describing the influence of the ISM on the NS. We assume that $\dot{M}$  is equal to the Bondi accretion rate (\cite{1944MNRAS.104..273B}):
\begin{equation}
    \label{M_dot}
    \Dot{M} = \beta\pi R_{\text{G}}^2 n m_{\text{p}} v_{\infty},   
\end{equation}
%Even if accretion is not possible $\dot M$ still can be used as the parameter characterizing the ISM interaction with the NS. 
where $m_{\text{p}}$ is a proton mass. Coefficient $\beta = 1-4$
depending on ratio between $v_{\infty}$ and $c_{\text{s}}$. If $c_{\text{s}}$ much smaller than $v_{\infty}$ (so that $v_{\infty} \approx v$) then $\beta \approx 1$. %In the case of a spherical accretion, it would be more correct to use the factor $4 \pi$ instead of $\pi$.

Following~\citet{1970SvA....14..527S}, the magnetized wind is capable of preventing the matter from being drawn in, if the wind pressure is larger than the pressure of the surrounding matter. We can find the radius $R_{\text{Sh}}$ (Shvartsman radius), which  
%in the 1D approximation 
is determined by the pressure balance between the wind and the surrounding material:

\begin{equation}
    \label{R_Sh}
    R_{\text{Sh}}=\left(  \frac{\xi \mu^2 (GM)^2 \omega^4}{\Dot{M} v^5 c^4} \right)^{1/2}.
\end{equation}
Here $R_{\text{Sh}} > R_{\text{G}}$ since equilibrium can only exist if the Shvartsman radius is larger than the gravitational capture radius (\cite{1970SvA....14..527S}). % is a citation necessary here?

While the NS spins down, the Shvartsman radius decreases until it equals either $R_{\text{l}}$ or $R_{\text{G}}$, depending on which radius is larger. In the latter case, the matter experiences near free-fall conditions before reaching $R_{\text{l}}$. After the matter enters the light cylinder, it begins to interact with the magnetic field. 
%The inequality $R_{\text{G}} > R_{\text{l}}$ can be rewritten in a form $P < 4\pi G M / (c v^2) \sim 2 \times 10^5 (v/(100\text{km/s}))^{-2}$~yrs. 
%We can estimate the time of reaching this period by integrating eq.~\ref{E_sd}. As a result, the time for the NS with a velocity $v \lesssim 100$~km/s to reach this period is much greater than the Galaxy lifetime, and 
%The condition $R_{\text{Sh}} \lesssim R_{\text{G}}$ must be fulfilled much earlier for all NSs that are being considered.
Typically, for INSs with reasonably low velocities $R_{\text{l}} < R_{\text{G}}$.
%For the parameters under consideration the condition $R_{\text{l}} < R_{\text{G}}$ is always fulfilled.
 Hence, the only period of ejector-propeller transition can be written from the condition $R_{\text{Sh}} = R_{\text{G}}$:

\begin{equation}
\begin{split}
\label{P_EP}
        P_{\text{EP}} = \frac{2\pi}{c} \left(\frac{\xi \mu^2}{4 \dot{M} v}\right)^{1/4}.  
        %260 \left(\frac{B}{10^{14}~\text{G}} \right)^{1/2} \times \\ \times \left(\frac{v}{100~\text{km/s}} \right)^{-1/4}\left(\frac{n}{0.3~\text{cm}^{-3}} \right)^{-1/4}~\text{s}
\end{split}
\end{equation}

For a `standard' NS with mass $1.4M_\odot$, moment of inertia $10^{45}$ g cm$^2$ and radius $R = 10$ km this formula can be rewritten as $P_{\text{EP}} \approx 190~B_{14}^{1/2} v_2^{1/2} n^{-1/4}$~s. Here, $B_{14}=B/(10^{14}~\text{G})$, $v_{2}=v/(100$~km~s$^{-1})$, $n = n/(1~\text{cm}^{-3})$ are assumed along with $\xi = 2$ (see Sec.~\ref{sect:model}). When NS rotation achieves this period, the ejector stage ends and the propeller stage begins. This estimate is based on the assumption that NS is an active radio pulsar, so that it spins down due to both vacuum and wind losses of rotational energy. However, if $P_\mathrm{EP} > P_\mathrm{death}$~--- a so-called 'death line' condition~--- then one has to consider that after crossing the death line only vacuum losses are relevant. Hence the effective magnetic moment $\mu_\mathrm{vac} = \mu\sin\alpha$ (e.g. \cite{2005AstL...31..263B}), which leads to
\begin{equation}
    P_\mathrm{EP,vac}=P_\mathrm{EP}\sqrt{\sin\alpha} < P_\mathrm{EP}.
\end{equation}
However, in the present work we are particularly interested in the fate of the long-period active radio pulsars, so we will omit this correction in the following. 

\subsection{Propeller}

The propeller stage is a necessary step between ejection and accretion. During this evolutionary stage, the interaction of the outer matter with the magnetic field prevents accretion onto the surface of the compact object. 
The balance between the magnetic field pressure and the pressure of the external matter determines the magnetosphere radius, $R_{\text{m}}$. In the simplest situation, it is equal to the so-called Alfv{\'e}n radius: 

\begin{equation}
    \label{R_A}
    R_A = \left( \frac{ \mu^2}{8 \Dot{M} \sqrt{2GM}}  \right)^{2/7}.
\end{equation}
A more detailed analysis of various situations can result in a different definition of $R_{\text{m}}$. E.g., according to \citet{1981MNRAS.196..209D}, $R_{\text{m}}$ slightly exceeds the Alfv{\'e}n radius due to the existence of an envelope consisting of the heated gravitationally captured material. In this case, $R_{\text{m}}$ represents the radius of the inner boundary of the envelope and can be written as follows:

\begin{equation}
    \label{R_m}
    R_{\text{m}} = R_A \left (\dfrac{R_{\text{G}}}{R_A} \right)^{2/9}.
\end{equation}

Since the magnetic field, which rotates rigidly with the NS, cannot exist outside of the light cylinder, we bound $R_{\text{m}}$ from above by the value of $R_{\text{l}}$.

In addition to $R_{\text{m}}$, we note the other important critical radius~--- the corotation radius $R_{\text{c}}$ at which the material can experience an equilibrium between gravitational attraction and centrifugal inertial force:
%If the material reaches the corotation radius $R_{\text{c}}$, which means that condition $R_{\text{m}} = R_{\text{c}}$ is fulfilled, in the very point it experiences an equilibrium between gravitational attraction and centrifugal inertial force,
%P^2}{4\pi^2}
\begin{equation}
    \label{R_c}
    R_{\text{c}} = \left( \frac{GM}{\omega^2}  \right)^{1/3}.
\end{equation}

During the propeller stage, a rapidly rotating magnetosphere prevents the material from going further than $R_{\text{m}}$. If the material remains at a radius larger than $R_{\text{c}}$ it cannot fall onto the NS because the centrifugal inertial force overrides the gravitational attraction. Therefore, we consider the propeller stage to exist as long as $R_{\text{m}} > R_{\text{c}}$.

%The deceleration of the NS at the propeller stage is due to the transfer of angular moment from the magnetosphere to the surrounding material. 
During the propeller stage, the rotational energy is dissipated at the inner boundary of an atmosphere. Then it is convected up and lost through the outer boundary. 
%We will rewrite the deceleration law $I \dot{\omega} = -K_{\text{P}}$ at the propeller stage in the following form:

%\begin{equation}
%    \label{P_sd}
%    \dv{P}{t}=\frac{P^2}{2\pi I} K_{\text{P}},
%\end{equation}
%where $K_{\text{P}}$ is the braking torque at the propeller stage. 

There are a number of ways to describe the NS deceleration at the propeller stage. Below we list some of the possible models proposed for the braking torque $K_{\text{P}}$ at this sage, starting with the one with the highest energy losses and ending with the one with the lowest.

\begin{enumerate}[label=\Alph*)]

\item \citet{1975SvAL....1..223S}. In this model, the external matter is considered to be thrown away at a speed close to the rotational speed at the magnetospheric boundary $\omega R_{\text{m}}$. So, the torque $K_{\text{P}}$ can be written as:

% Переписать в безразмерном виде через r_{\text{m}} = R_{\text{m}} / R_{\text{G}} и r_{\text{c}} = R_{\text{c}} / R_{\text{G}}
\begin{equation}
    \label{K_P1}
        K_{\text{P}}=\Dot{M} \omega R_{\text{m}}^2 = 
        %\frac{1}{8\sqrt{2}}\left(\frac{R_{\text{G}} R_{\text{m}}^{1/2}}%{R_{\text{c}}^{3/2}}\right)\frac{\mu^2}{R_{\text{m}}^3}.
        \frac{1}{8\sqrt{2}}r_{\text{c}}^{-3/2}r_{\text{m}}^{1/2}\frac{\mu^2}{R_{\text{m}}^3},
\end{equation}
where $r_{\text{c}} = R_{\text{c}}/R_{\text{G}}$ and $r_{\text{m}} = R_{\text{m}}/R_{\text{G}}$ are dimensionless radii.

% Второй график и вторая часть таблицы пересчитаны. В одном месте поправила текст результатов там, где он ссылается на второй график, больше мест пока не нашла.
\item \citet{1973ApJ...179..585D}. Here it is assumed that the material carries away the angular momentum at the free fall velocity $v_{\text{ff}} = \sqrt{2GM/R_{\text{m}}}$, hence
%Here it is assumed that the material carries away the Keplerian mechanical moment, hence
\begin{equation}
\label{K_P2}
% новая формула
    K_{\text{P}} = \dot{M} \sqrt{2GMR_{\text{m}}} = %\frac{1}{8} \left(\frac{R_{\text{G}}}{R_{\text{m}}}\right) 
    \frac{1}{8} r_{\text{m}}^{-1}
    \frac{\mu^2}{R_{\text{m}}^3}.
\end{equation}

\item \citet{1975A&A....39..185I}. %The material is thrown back at the free fall velocity $v_{\text{ff}} = \sqrt{2GM/R_{\text{m}}}$. 
Within this model, the assumed spin-down law is $I \omega \dot{\omega} = - \dot{M} v_{\text{ff}}^2 / 2 = - GM\dot{M} /R_{\text{m}}$ and the corresponding braking torque is equal to

\begin{equation}
    \label{K_P3}
    K_{\text{P}} =  \frac{GM\dot{M}}{\omega R_{\text{m}}} = %\frac{1}{8\sqrt{2}}\left(\frac{R_{\text{G}} R_{\text{c}}^{3/2}}{R_{\text{m}}^{5/2}}\right)\frac{\mu^2}{R_{\text{m}}^3}.
    \frac{1}{8\sqrt{2}}r_{\text{c}}^{3/2}r_{\text{m}}^{-5/2}\frac{\mu^2}{R_{\text{m}}^3}.
\end{equation}

\item The supersonic propeller considered by \citet{1979MNRAS.186..779D} and \citet{1981MNRAS.196..209D}. In this approach, the rotational losses are considered to be constant $I \omega \dot{\omega} = - \dot{M} v^2 / 2$, where $v$ contains spatial velocity $v_{\infty}$ and the acoustic speed $c_{\text{s}}$: $v = \sqrt{v_{\infty}^2 + c_{\text{s}}^2}$. From this, we can obtain the braking torque:
\begin{equation}
    \label{K_P4}
        K_{\text{P}}= \frac{1}{8\sqrt{2}} r_{\text{c}}^{3/2} r_{\text{m}}^{-3/2}\frac{\mu^2}{R_{\text{m}}^3}.
\end{equation}

\end{enumerate}

% Возможное место для объяснения, почему мы рисуем первые два варианта пропеллера (с большими энергопотерями), а вторые два не рисуем (потери меньше пульсарных). Энергопотери пропеллера должны быть больше энергопотерь эжектора (но почему обязательно так? Почему взаимодействие с веществом должно забирать больше энергии, чем НС отдает на стадии эжектора?).

% We assume that some of the material can penetrate the magnetosphere if $R_{\text{m}} < R_{\text{c}}$. 
Once the material reaches a radius smaller than the corotation radius, the centrifugal barrier will no longer prevent it from falling onto the NS. The corresponding transition period 
can be then calculated from the condition $R_{\text{c}} = R_{\text{m}}$ as

%Therefore, $R_{\text{m}} < R_{\text{c}}$ is the necessary condition for the transition from propeller to accretor. The characteristic period for $R_{\text{c}} = R_{\text{m}}$ is

%\begin{equation}
 %   P_{\text{PA}} = \frac{2\pi\mu^{2/3}}{2^{5/6}(GM)^{1/3}\dot{M^{1/3}}v^{2/3}}
%\end{equation}

\begin{equation}
\label{P_PA}
    P_{\text{PA}} = \pi \left(\frac{\mu^2 \sqrt{2}}{GM\dot{M}v^2}\right)^{1/3}.
\end{equation}
Using the same parameters of a 'standard' NS this formula can be written in a way similar to $P_{\text{EP}}$ as $P_{\text{PA}} = 3 \times 10^5 B_{14}^{2/3} v_2^{1/3} n^{-1/3}$~s. If the period of the NS exceeds $P_{\text{PA}}$, we consider the NS to change the stage from propeller to accretor.

\subsection{Accretor}
% Тут я много раз цитирую Шакуру, Постнова для settling accretion, но их статей по этой теме очень много и все похожи. Как тут правильно цитировать серию статей?

% Результаты расчётов с аккрецией с оседанем будут в разделе с результатами, а не в дискуссии.
Since accretion starts, several different regimes can be realized depending on circumstances. Following \citet{2012MNRAS.420..216S} we distinguish supersonic (i.e. standard, Bondi or Bondi-Hoyle-Littleton) accretion and subsonic settling accretion. 
% \cite{2017arXiv170203393S}

The first regime is spherical (or cylindrical) accretion of free-falling matter with an accretion rate $\dot{M}$. The most important spatial scale here is the corotation radius $R_c$ (see e.g., the derivation of the spin-down torque by \cite{1992ans..book.....L}). 
So, neglecting angular momentum accreted from ISM, the spin-down law is $I \dot{\omega} = -K_{\text{A}}$, where the angular momentum $K_{\text{A}}$ can be expressed as follows:
\begin{equation}
    \label{K_A}
    K_{\text{A}}=\zeta\frac{\mu^2}{R_{\text{c}}^3}.
\end{equation}
Here $\zeta$ is a dimensionless coefficient of the order of unity. It is usually assumed to be $\zeta \sim 1/3$ (\cite{1992ans..book.....L}). 

%At this stage, the magnetosphere is surrounded by a hot envelope, so normal accretion cannot proceed. 

However, the Bondi accretion only occurs if the surrounding material is effectively cooled. A more correct description of the accretion regime realized in our case is the subsonic settling accretion. It is characterized by the existence of a quasi-static atmosphere extending from $R_{\text{m}}$ to $R_{\text{G}}$. This heated plasma surrounding the NS cools due to convective motions and bremsstrahlung radiation. 
%In this case, the energy input to the atmosphere due to the rotation of the magnetosphere dominates the radiative losses. 
Thus the characteristic cooling time $t_{\text{cool}}$ is significantly longer than the free-fall time $t_{\text{ff}}$. This prevents standard accretion with an accretion rate of $\dot{M}$ defined by eq.~(\ref{M_dot}). 

As it is shown by~\citet{2015MNRAS.447.2817P}, at the settling accretion phase the spin-down torque $K_{\text{SA}}$ is slightly different from eq.~(\ref{K_A}). Namely:
%$\dot{P} = {8 \dot{M} R_{\text{m}}^2 P}/{I}$. If we rewrite it in a form $I\dot{\omega}=-K_{\text{SA}}$, then the braking torque at the settling accretion stage is
    %K_{\text{SA}} = 8 \dot{M} \omega R_{\text{m}}^2 = \frac{1}{\sqrt{2}}r_c^{-3/2}r_m^{1/2}\frac{\mu^2}{R_{\text{m}}^3},

\begin{equation}
    \label{K_SA}
    K_{\text{SA}} = 8 \dot{M} \omega R_{\text{m}}^2 = \frac{1}{\sqrt{2}}r_c^{3/2}r_m^{-5/2}\frac{\mu^2}{R_{\text{c}}^3},
\end{equation}
which coincides with the propeller spin-down torque up to a factor of 8.

According to \citet{2015ARep...59..645S}, the matter in the envelope slowly flows towards the NS with an average velocity $u = (t_{\text{ff}} / t_{\text{cool}})^{1/3} v_{\text{ff}}$, where $v_{\text{ff}}$ is a free-fall velocity. So the maximum possible accretion rate is $\dot{M}_{\text{SA}} \sim (t_{\text{ff}} / t_{\text{cool}})^{1/3} \dot{M}$. Since $\dot{M}_{\text{SA}} \ll \dot{M}$, the X-ray luminosity of potential observable sources could be a few orders of magnitude lower than standard estimates, which take $\dot{M}$ to be an accretion rate.

%The heated envelope can only exist if $t_{\text{ff}} / t_{\text{cool}} \lesssim 1$. \cite{1981MNRAS.196..209D} have shown that the condition for the \\ bremsstrahlung cooling to become important $t_{\text{ff}} \sim t_{\text{cool}}$ leads to the collapse of the envelope, and it can be rewritten in a form of reaching a period $P = P_{br}$. The period $P_{br}$, as it also discussed by \cite{2001A&A...368L...5I}, can be calculated as $P_{br} \simeq 3 \times 10^{8} B_{14}^{16/21} v_2^{15/7} n^{-5/7}$~s. We assume that this condition corresponds to transition to the standard accretion regime. As long as $P > P_{br}$, the Bondi accretion begins. 

Considering the settling accretion regime, \citet{2012MNRAS.420..216S} have shown that the hot envelope surrounding the NS can be also effectively cooled by the Compton process. However, there is a critical value of the X-ray luminosity at which the Compton cooling time becomes equal to the free-fall time. This condition corresponds to $\dot{M} \sim 10^{16}$~g~s$^{-1}$. This accretion rate is much larger than the possible accretion rate from the ISM. So it can be assumed that the heated material cools only due to bremsstrahlung radiation and the settling accretion regime is established.

On the other hand, \citet{2002A&A...381.1000P} have considered the late stages of the Bondi accretion. It is shown that during the standard accretor stage the evolution of the NS period can be influenced by the turbulent angular momentum from the ISM, since the ISM material directly interacts with the magnetosphere. It can both accelerate and decelerate rotation of the NS. The influence of turbulence starts to be important when the NS spins down to the period $P_{\text{cr}}$:
%One can write the characteristic timescale of the turbulent influence $\tau_{turb}$ as well as the timescale of a spin-down process $\tau_{mag}$ at the Bondi accretion eq.~(\ref{K_A}). The critical period for the timescales to become of the same order is

\begin{equation}
\label{P_cr}
    P_{\text{cr}} = \frac{2\pi \mu \sqrt{\zeta}}{\sqrt{GM\dot{M}j}},
\end{equation}
where specific angular momentum $j$ can be expressed through the characteristic turbulent velocity $v_t\simeq 10~\text{km}~\text{s}^{-1}$ at the scale $R_t\simeq2\times10^{20}$~cm: $j = v_t R_t^{-1/3} R_{\text{G}}^{4/3}$. 
$P_{\text{cr}}$ is defined by $K_{\text{A}}=j \dot{M}$. %проверить!

If $j$ exceeds the Keplerian moment at the magnetosphere radius $j_K = \sqrt{GMR_{\text{m}}}$, the disk is formed and the angular momentum in eq.~(\ref{P_cr}) is assumed to be $j_K$ instead of $j$. However, the turbulent angular momentum of the ISM for our parameters $j < j_K$, so we can rewrite eq.~(\ref{P_cr}) in a form $P_{\text{cr}} = 10^7 B_{14}v_2^{17/6}n^{-1/2}$~s.

When $P \sim P_{\text{cr}}$ the two processes become comparable. For the periods $P \gtrsim P_{\text{cr}}$ the NS period is significantly affected by the turbulence during the deceleration. 

According to the consideration of the further influence of the turbulent angular momentum on the deceleration of the NS, conducted by \citet{2001astro.ph.10022P}, the NS reaches `spin equilibrium'. The period evolution begins to be completely determined by the turbulent forces. It fluctuates around the turbulent period: 

\begin{equation}
\label{P_turb}
    P_{\text{turb}} = 8 \times 10^9 B_{14}^{2/3} v_2^{43/9} n^{-2/3}~\text{s}
\end{equation}

After reaching the turbulent period the NS can both accelerate and decelerate with a characteristic timescale $\sim R_{\text{G}} / v$. % Перенесено в модель: However, in our calculations we will assume the period to be constant.

%Это неверно: During the settling accretion the NS can also be influenced by the ISM turbulence. Similar to the case with the standard accretion (eq.~\ref{P_cr}), we write down the critical period for the characteristic time of turbulent spin-up/spin-down to be of the same order as deceleration described by the eq.~(\ref{K_SA}):
% Я бы лучше упомянула не равенство хар. времен изменения P, а время охлаждения. Потому что пока вокруг НЗ горячая оболочка, внешний момент напрямую взаимодействовать с магнитосферой не будет

%\begin{equation}
%    P_{\text{cr}}' = \frac{16 \pi R_{\text{m}}^2}{j}
%\end{equation}

%Для пропеллера Шакуры, v = 30 км/с, модели затухания поля ED и для B0 = 10^14 Гс $P_{cr1} > P_{\text{turb}}$, что не имеет физ. смысла 

\section{Model}
\label{sect:model}
%\cite{2014MNRAS.441.1879P} have shown that in this case the braking torque can be described by the following equation:
%\dv{}{t} \left( \frac{I\omega^2}{2} \right)
%\begin{equation}
%    \label{E_sd_old}
    %I\dv{\omega}{t}
%     K_{\text{E}}=-\frac{\mu^2 \omega^3}{c^3} \left(k_0 + k_1\sin^2\alpha\right),
%\end{equation}
%where $\alpha$ is the magnetic angle between the spin and the magnetic moment, $k_0, k_1$ are dimensionless coefficients.
%Taking the evolution of the magnetic angle~$\alpha$ into account requires a less studied distribution of NSs over the magnetic angles. At the same time, it does not make a decisive contribution to the final result due to the small factor $k_1 \approx 1.4$. Therefore we neglect the dependence on the angle in formula~\ref{E_eq_old}. Then it can be rewritten as follows:
In our work we focus on the modeling $P(t)$ curves and deriving the moments of time when an NS starts to accrete: $t=t_{\text{PA}}$ (when $P=P_{\text{PA}}$), enters the turbulent regime: $t = t_{\text{cr}}$ (when $P=P_{\text{cr}}$), and reaches the equilibrium: $t=t_{\text{turb}}$ at the period $P_{\text{turb}}$. 

We consider the spin period evolution of an INS with a given initial period $P_0$, initial equatorial magnetic field $B_0$, and velocity $v_\infty$ through the ISM of number density $n$. Within the calculations we assume $P_0 = 100$~s; $B_0 = 10^{11}$, $10^{12}, 10^{13}$ or $10^{14}$~G; $v_\infty = 30$ or $100$ km~s$^{-1}$ and $n = 0.3$~cm$^{-3}$. Also we try two models of propeller spin-down torque (see the text below) and three models of the magnetic field decay (see Sec.~\ref{sect:field_decay}). Finally, we
assume $v_\infty \gg c_{\text{s}}$, since sound velocity in the ISM $c_\text{s} \sim \sqrt{kT/m_\text{p}} \sim 10$~km~s$^{-1}$ for $T \sim 10^4$~K (\cite{2016SAAS...43...85K}). So, in our calculations $v = \sqrt{v_\infty^2 + c_{\text{s}}^2} \approx v_\infty$.

%In our calculations, an NS has a set of parameters: initial period $P_0$ and magnetic field strength $B_0$, the number density of the ISM $n$, and the velocity $v$. We limit the duration of each track to the Galaxy lifetime $t_{\text{gal}} = 1.35 \times 10^{10}$~years. %since the studied phenomena have a timescale much greater than $t_0$. %start our calculations with the time $t_0 = 1$~year and 
% the field decay model (CF, ED or HA), 

While the NS spins down, it changes its evolutionary stages. To determine the initial evolutionary stage, the value of $P_0$ compares to transition periods of between ejector and propeller $P_{\text{EP}}$ (eq.~\ref{P_EP}) and propeller-accretor $P_{\text{PA}}$ (eq.~\ref{P_PA}). 

At each stage we adopt the spin-down law in the from:
\begin{equation}
    \dv{P}{t}=\frac{P^2}{2\pi I} K,
\end{equation}
%$\dot{P} = f(t, P, B, n, v)$.
where the spin-down torque $K$ is
 $$
    \label{K}
    K =
    \left\{
    \begin {array} {ll}
    \displaystyle
    K_{\text{E}}
    , & P \le P_{\text{EP}} \\
    \displaystyle
    K_{\text{P}}
    , & P_{\text{EP}} < P \le P_{\text{PA}} \\
    K_{\text{A}}
    , & P_{\text{PA}} < P \le P_{\text{turb}} \\
    \end {array}
    \right.
$$

The particular form of a spin-down torque depends on an evolutionary stage:
\begin{itemize}
    \item Ejector. For the ejector stage we take $K_{\text{E}}$ in a form of eq.~(\ref{K_E}). In our paper, we consider NSs with long periods $\gtrsim 100$~s and relatively low velocities $v \lesssim 100$~km~s$^{-1}$. %(30~km~s$^{-1}$ and 100~km~s$^{-1}$). v_{\infty}
    For periods and magnetic field values necessary for those NS to be ejectors (>$10^{13}$~G) the magnetic alignment timescale is longer than the Galaxy lifetime. Therefore we neglect the change in the angle $\alpha$ over time. Assuming a uniform distribution of newborn NSs over angles we use the mean value of $\langle \sin^2\alpha \rangle = 2/3$. So, the factor $\xi = k_0 + k_1\sin^2\alpha \approx 1.93$ and we take $\xi = 2$ exactly. %Following \cite{2014MNRAS.441.1879P} we use $k_0 \approx 1, k_1 \approx 1.4$. %this period value

    \item Propeller. Hereafter we consider only two variants of $K_{\text{P}}$: case A (eq.~\ref{K_P1}) and case B (eq.~\ref{K_P2}). We do not use C and D models, since the rotational losses in cases C and D can be smaller than the pulsar losses. %This will cause the ejector stage to become shorter than the propeller stage, which 
    This contradicts the conclusions by \citet{1995AZh....72..711L} who demonstrated that spin-down at the ejector stage might be less efficient that at the propeller stage.
    %, where the duration of the ejector stage is shown to be longer. Therefore, only the two first models of the propeller spin-down are used in the calculations below. 

    \item Accretor. We assume the accretion to be standard and use the spin-down torque $K_{\text{A}}$ in a form of eq.~(\ref{K_A}) until the equilibrium is reached ($P = P_{\text{turb}}$).

\end{itemize}

After adding the initial condition $P(t_0) = P_0$ the initial value problem is numerically solved until the time limit is reached or the period is equal to the transition period (for the ejector and propeller stages this period is respectively $P_{\text{EP}}$ and $P_{\text{PA}}$). In the latter case, the new initial value problem is solved. It contains the spin-down law according to a new evolutionary stage and the initial condition in a form $P(t_1) = P_1$, where $P_1$ is the corresponding transition period, which was reached during previous calculations, and $t_1$ is the time of the transition to this stage. 

At the accretor stage, we are also interested in the moment of NS entry into the turbulent regime $t_{\text{cr}}$ for $P=P_{\text{cr}}$ ($P_{\text{cr}}$ is described by eq.~\ref{P_cr}), where the turbulence can significantly influence the NS period. For $t > t_{\text{cr}}$ the period evolution cannot be described only by the deceleration law with the spin-down torque from eq.~(\ref{K_A}), because of the existence of the stochastic turbulent angular momentum. However, we can give a rough upper bound for the period in this regime by assuming that the deceleration law remains the same. So, the period is calculated by solving the corresponding initial value problem until $P = P_{\text{turb}}$ (eq.~\ref{P_turb}) is reached at $t_{\text{turb}}$. After $t_{\text{turb}}$ the spin period is assumed to fluctuate around the constant value $P_{\text{turb}}$. Since we do not model the evolution of the existing specific objects, we will only show the mean period value $P_{\text{turb}}$ at this regime. So, after the NS reaches the accretor stage the period obeys the corresponding deceleration law. If the accretor stage starts with the period  $P=P_{\text{PA}}>P_{\text{cr}}$, then we assume that the turbulent regime is already reached at $t_{\text{PA}}$.

%We mark the exact moment when turbulence become significant ($P=P_{\text{cr}}$), and if the accretor stage starts with the period $P=P_{\text{PA}}>P_{\text{cr}}$, this moment will correspond to the transition to the turbulent regime ($P=P_{\text{PA}}$ and $t=t_{\text{PA}}$).

%Then taking into account the factor $\xi = \left(k_0 + k_1\sin^2\alpha\right) \approx 2$ and the spin-down moment at the ejector stage $K_{\text{E}} \equiv -I \dot{\omega}$, the eq.~(\ref{E_sd}) can be rewritten as follows:
%The spin-down torque in the spin-down law we use for each stage are: eq.~(\ref{K_E}) at the ejector stage, for the propeller stage the spin-down torques described by eq.~(\ref{K_P1}) for the case A and eq.~(\ref{K_P2}) for the case B, for the accretor deceleration we use the spin-down torque in a form of eq.~(\ref{K_A}).

%The transition periods are: $P_{\text{EP}}$ (eq.\ref{P_EP}) for the ejector-propeller and $P_{\text{PA}}$ (eq.~\ref{P_PA}) for the propeller-accretor transition.

\subsection{Magnetic field evolution}
\label{sect:field_decay}

Magnetic field evolution is a very important ingredient of our modeling. 
Particularly, we are interested in magnetic field decay. 
Presently (see a review by \cite{2021Univ....7..351I}), field decay is much better understood for young INS with ages $\lesssim 10^6$~yrs. 

%velocity. V_min for PSR 0901-4846 and others. 
%NSs with the same period, field, age, and in the same ISM can be at different stages if they have different velocities. We assume that long-period NSs can have different velocities. They can be visible as radio pulsars only if their velocities are sufficiently high.

Within our work, we consider three models of field behavior:

\begin{enumerate}

\item Model CF: constant field. For illustrative purposes, we perform calculations with a constant field. Despite this assumption is not realistic, it helps for a better understanding of important aspects of magneto-rotational evolution focusing on spin properties. 

\item Model ED: continuous exponential field decay. 
Quite often it is assumed that on a long time scale, the field decays exponentially with the same rate. E.g., this can be due to the Ohmic decay due to impurities.
 Then, the field evolution is described with a very simple equation:

\begin{equation}
  B = B_0 \exp(-t/\tau_{\text{Ohm}}).
\end{equation}

To derive an estimate for $\tau_{\text{Ohm}}$ we use the following consideration. Let us consider the initial field to be typical for normal radio pulsars: $B=10^{12}$~G. Then we require that on the time scale of the order of the Galactic age the field drops to a value typical to millisecond pulsars: $10^8$~G.
% due to Ohmic decay after the lifetime of the Galaxy~$13.6\times 10^{9}$~years.
 Then we obtain the value ~$\tau_{\text{Ohm}} = 1.48 \times  10^{9}$~years which is used in our calculations.

\item Model HA: Hall attractor. In this model we apply a rapid initial field evolution due to the joint influence of the Ohmic decay and Hall cascade (see e.g., 
\cite{2019LRCA....5....3P} and references therein). 

Field decay in this scenario is calculated according to the equation proposed by \citet{2008A&A...486..255A}:

\begin{equation}
    \label{B}
    B=B_0\frac{\exp(-t/\tau_{\text{Ohm}})}{1+(\tau_{\text{Ohm}}/\tau_{\text{Hall}})(1-\exp(-t/\tau_{\text{Ohm}}))}.
\end{equation}
Here $B_0=\mu/R^3_{\text{NS}}$ is the initial value of magnetic field strength on the NS equator, $\tau_{\text{Ohm}}\sim 10^6$~yrs is Ohmic decay characteristic time, $\tau_{\text{Hall}} \sim10^4/(B_0/10^{15}\text{~G})$~yrs is Hall cascade characteristic time scale corresponding to the initial magnetic field.

The rapid initial field decay due to the Hall cascade is terminated when the magnetic field configuration reaches the so-called Hall attractor. 
The existence of this stage is firstly considered by~\citet{2014PhRvL.112q1101G}. These authors demonstrated that on the time scale of few~$\times\tau_{\text{Hall}}$  the Hall cascade saturates. We assume that the magnetic field decays according to eq.~(\ref{B}) until it reaches the value~$B\approx 0.05B_0$ (i.e., approximately three $e$-foldings, see \cite{2014MNRAS.438.1618G}). Afterwards, the field is assumed to be constant. 

\end{enumerate}

\section{Results}

First, let us estimate the time for the NS to reach the accretor stage. We integrate eq.~(\ref{Euler}) for the spin period at the propeller stage using $K_\text{E}$ in the form of eq.~(\ref{K_E}) assuming the magnetic field is constant. Therefore, the time required for the NS to spin down from $P_0$ to the transition period $P_\text{EP}$ is:

% Из всего набора параметров таблицы 1 только для НЗ с полем 10^11 Гс и скоростью 100 км/c можно получить P_min (=6.49e+00 s), для остальных P_min из t_E + t_P = t_gal не получится.

\begin{equation}
    \label{t_E}
    t_\text{E} = \frac{Ic^3}{8\pi^2\xi \mu^2}\left(P_\text{EP}^2 - P_0^2\right).
\end{equation}

If $P_\text{EP} \gg P_0$, then $t_\text{E} \approx Ic / (4\mu\sqrt{\xi\dot{M}v}) \approx$ \\ $\approx 2\times 10^7 B_{14}^{-1}v_2 n^{-1/2}$~yrs.

%Icv / (8GM\mu\sqrt{\xi\pi n m_\text{p}})$ Я оставлю только формулу 24 или 26

To obtain the duration of the propeller stage we integrate eq.~(\ref{Euler}) using $K_\text{P}$ in the form of eq.~(\ref{K_P1}) for case A and eq.~(\ref{K_P2}) for case B. Thus, the duration of the propeller stage can be expressed as follows:

% R_m > R_l в начале стадии пропеллера. Если вставить R_m = min(R_m, R_l) в формулы для момента и посчитать численно, то длительность пропеллера будет незначительно отличаться. Так, графики tB100 и tB100CF отличаются именно из-за этого. Поэтому, если делать формулу t_P абсолютно точной, пропеллер нужно разбить на две части, где в первой части в выражении для t_P будет R_l вместо R_m. Граничный период между частями пропеллера это P(R_m = R_l) = 2 pi R_m/c 
\begin{equation}
    \label{t_P}
    t_\text{P} \approx 
    \left\{
    \begin {array} {ll}
    \displaystyle
    \frac{I}{\dot{M}R_\text{m}^2} \ln{\frac{P_\text{PA}}{P}} %{P_\text{EP}}}
    % = \frac{I}{12\dot{M}R_\text{m}^2} \ln{\frac{c^{12}\mu^{2}}{2^{4}\xi^{3}\dot{M}(GM)^{4}v^{5}}}
    , & \text{case A} \\
    
    %\ln{\frac{c\mu^{1/6}}{2^{1/3}\xi^{1/4}\dot{M}^{1/12}(GM)^{1/3}v^{5/12}}}, & \text{case A} \\
    \displaystyle
    \frac{2\pi I}{\dot{M}\sqrt{2GMR_\text{m}}}\left(\frac{1}{P}-\frac{1}{P_\text{PA}}\right), & \text{case B} \\
    %P_\text{EP}
    \end {array}
    \right.
\end{equation}
Here $P$ represents the initial period at the propeller stage, so $P \ge P_\text{EP}$.

\begin{figure}
%\centering
\includegraphics[width=0.99\linewidth]{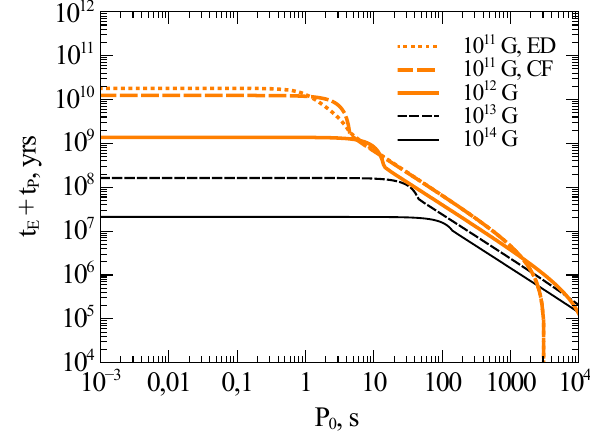}
\caption{The time for the NS to reach the accretor stage, which is assumed to be the total duration of the ejector and propeller stages $t_E+t_P$, versus the initial spin period $P_0$. The results are shown for case B of propeller spin down, $v=30$~km/s, $n=0.3$~cm$^{-3}$. We show two decay models only for $10^{11}$~G, since for the larger magnetic field values these two models lead to almost the same results of $t_E+t_P$.}
\label{tP}
\end{figure}

We numerically calculate several scenarios to obtain the dependence of the time required for the NS to become an accretor ($t_\text{E} + t_\text{P}$) on the initial spin period $P_0$. These scenarios include NSs with different magnetic field values and are shown in Fig.~\ref{tP}. To make our calculations more precise we choose only the lowest value of velocity, $v=30$~km~s$^{-1}$, assuming that low-velocity NSs have the highest chance to remain in the Galaxy disk, where the number density is $n\approx0.3$~cm$^{-3}$ and remains approximately constant. For illustrative purposes, we calculate only case B of propeller spin down %, since the duration of the propeller stage in case B is comparable to the ejector stage duration, 
 and show only field decay models CF and ED. 

Now, we will examine the obtained results. In most cases $t_\text{E} > t_\text{P}$, hence the total duration $t_\text{E} + t_\text{P}$ of NSs with low initial spin periods depends mostly on $t_\text{E}$. The total duration remains almost the same until $P_0$ becomes comparable to $P_\text{EP}$, since $t_\text{E}$ has a weak dependence on $P_0$ if $P_0 \ll P_\text{EP}$ (eq.~\ref{t_E}). As $P_0$ approaches $P_\text{EP}$, the duration of the ejector stage decreases and at some point becomes comparable to the duration of the propeller stage. As soon as $t_\text{E}$ and $t_\text{P}$ are equal, we see a break in the curves. For greater values of $P_0$ the total duration is determined by $t_\text{P}$ and has a stronger dependence on the initial spin period. If $P_0 > P_\text{EP}$, the NS starts its evolution from the propeller stage and $t_\text{E} = 0$. After $P_0 > P_\text{PA}$, the total duration is assumed to be zero, since the NS starts as an accretor. This situation is realized for $B_0 = 10^{11}$~G , $P_0 \gtrsim 3000$~s. Generally, the duration of both the ejector and propeller stages decreases with the increase in $B_0$. As for magnetic field decay, it changes the result only if the total duration is noticeably greater than the characteristic decay timescale. This is realised for $10^{11}$~G, $P_0 \lesssim 10$~s, since $t_\text{E} + t_\text{P} > \tau_\text{Ohm} \sim 10^9$~years. For $B_0 > 10^{11}$~G the field decay effect is negligible for the models that we analyze. 

% Можно сделать короткий вывод про необходимость большого периода здесь или вписать его в заключение

%In addition, we can express the initial period through the duration of the ejector stage $t_\text{E}$ and the final period that can be reached at this stage, that we assume to be the transition period $P_\text{EP}$.
%\begin{equation}
 %   \label{P_min}
  %  P_\text{min} = \sqrt{P_\text{EP}^2 - \frac{8\pi^2 \xi\mu^2 t_\text{E}}{Ic^3}}.
%\end{equation}
%where $t_\text{E}$ is the ejector stage duration.
%If we assume the duration of both the ejector and the propeller stages $t_\text{E} + t_\text{P}$ to be equal to the Galaxy age, then the obtained initial period $P_\text{min}$ will express the minimum initial period of the NS to reach the accretor stage within the Galaxy lifetime.

\begin{figure*}
\centering
\includegraphics[width=1\linewidth]{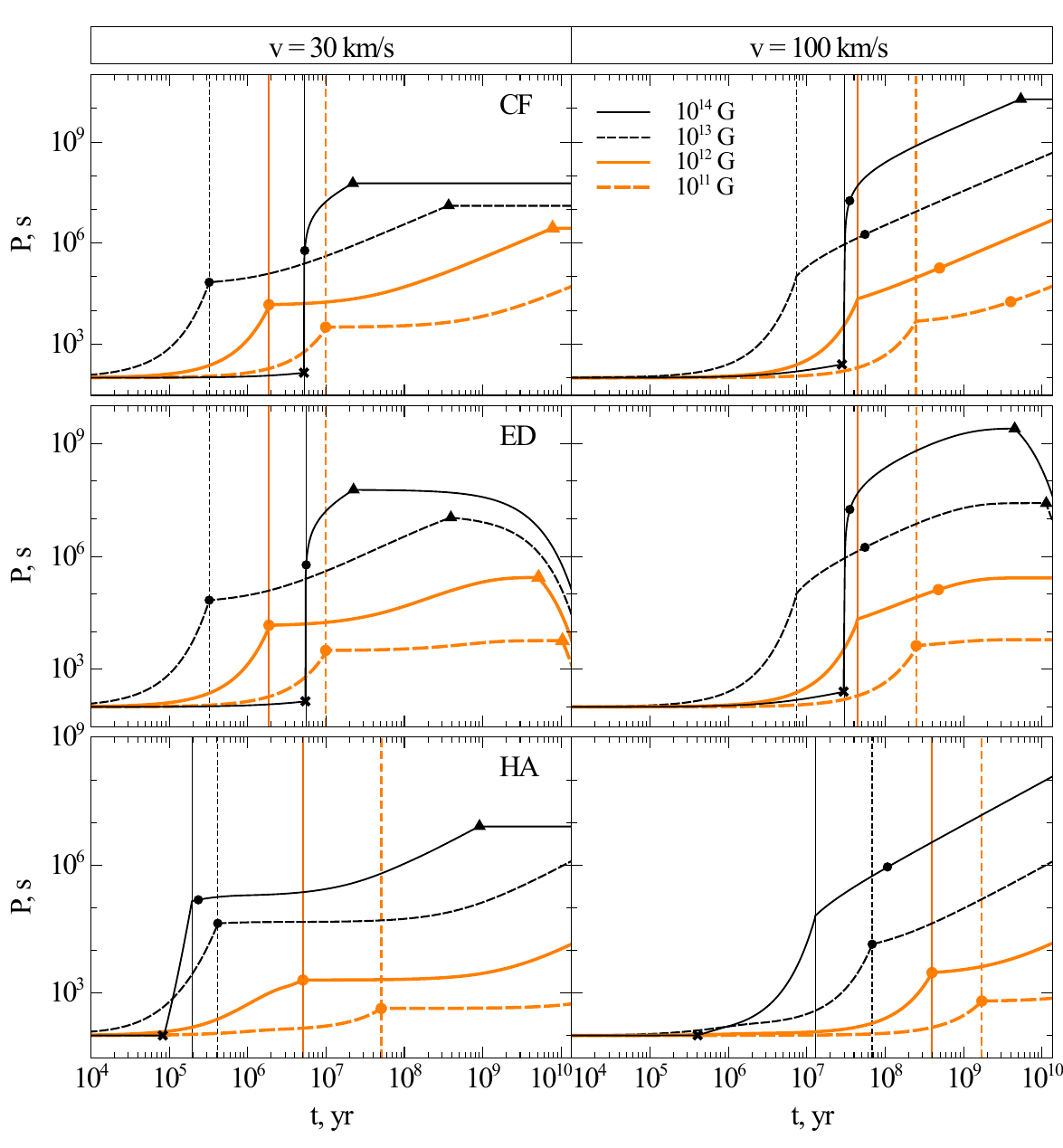}
\caption{The spin period evolution for the case A of propeller spin-down (eq.~\ref{K_P1}). %, where the spin-down moment is $K_{\text{P}}=\Dot{M} \omega R_{\text{m}}^2$. 
Two columns correspond to different values of the velocity $v$. Three rows represent different magnetic field evolution models: constant field (CF)~--- top, exponential decay (ED)~--- middle, and Hall attractor (HA)~--- bottom. Each line style corresponds to an initial magnetic field value $B_0$. %: $10^{11}, 10^{12}, 10^{13}$, and $10^{14}$~G.
Vertical lines correspond to the onset of accretion at $t_{\text{PA}}$ for every curve. An ejector-propeller transition of NS with $B_0 = 10^{14}$~G is seen as a cross. A filled circle on each curve is for the time $t_{\text{cr}}$ and the period $P_{\text{cr}}$. After the time $t_{\text{cr}}$ is reached the NS enters the turbulent regime and can also reach the turbulent period $P_{\text{turb}}$ at $t=t_{\text{turb}}$, which corresponds to a triangle.}
\label{ShakuraSd}
\end{figure*}

\begin{figure*}
\centering
\includegraphics[width=1\linewidth]{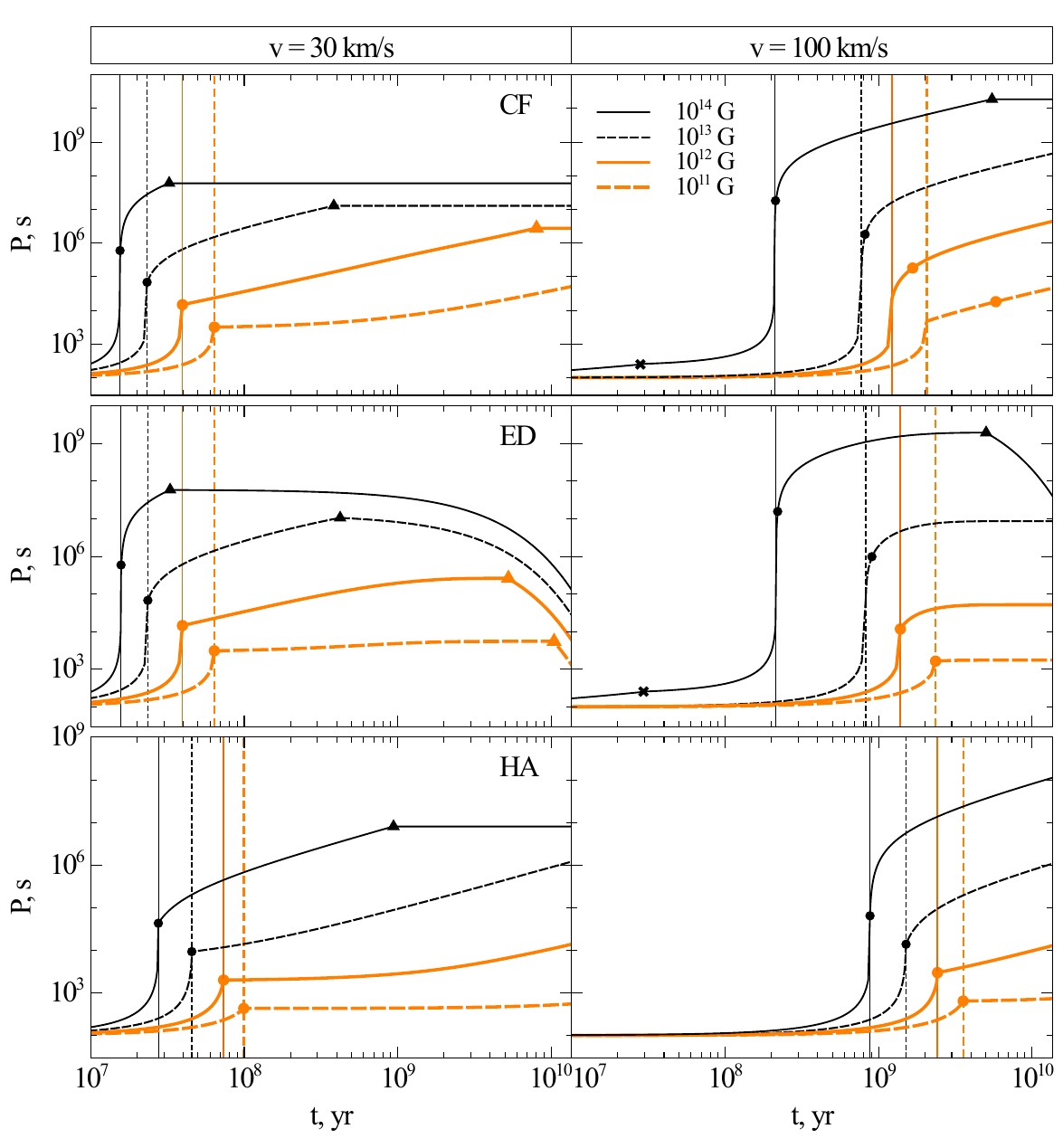}
\caption{The spin period evolution for the case B of propeller spin-down (eq.~\ref{K_P2}). An ejector-propeller transition of NS with $B_0 = 10^{14}$~G is shown as a cross. Vertical lines show the onset of accretion for each NS ($t_{\text{PA}}$). Filled circles correspond to the start of a turbulent regime ($P_{\text{cr}}$, $t_{\text{cr}}$). The reaching of $P_{\text{turb}}$ is shown as filled triangles. Line styles and colors are the same as in Fig.~\ref{ShakuraSd}.}
%D\&O spin-down, $R_{\text{m}} \sim R_A^{7/9}$}
\label{DOsd}
\end{figure*}

To follow the evolution of the NS rotation in more detail, we model 48 scenarios of the evolution of the INS spin-down: 2 variants of the propeller spin-down $\times$ 2 values of the velocity $v$ $\times$ 3 models of the magnetic field decay $\times$ 4 values of the initial magnetic field. All of them are presented in Fig.~\ref{ShakuraSd} and Fig.~\ref{DOsd} for the propeller models A and B, correspondingly.

At each plot, we show $P(t)$ curves for different initial magnetic fields $B_0$ assuming one of the models of the magnetic field decay~--- either constant field (CF) or exponential decay (ED) or Hall attractor (HA). For each curve, we mark moments of time $t_\text{PA}$ of the propeller-accretor transition by the vertical line of a corresponding style. Also, we use filled circles to mark moments of time $t_\text{cr}$ when the turbulence of ISM becomes important. Triangles indicate the reaching of an equilibrium for each NS. The ejector-propeller transition is marked with crosses. For the constant magnetic field model, we summarize all these times (and corresponding periods) in Table~\ref{tab:tab}. 

Generally, we conclude, that an INS with parameters under study would become an accretor in $\sim 10^5-10^9$ years of evolution.

%In this section, we present the results of our modeling of the spin evolution of INSs with long initial spin periods. 

% and accretor models (only Bondi accretion or settling accretion with the transition to Bondi accretion)
%We model the period evolution of the NS with an initial period $P_0 = 100$~s, initial magnetic field strength at the equator $B_0 = 10^{11}, 10^{12}, 10^{13}$, and $10^{14}$~G in an ISM with a number density $n = 0.3$~cm$^{-3}$. For the each of two propeller models (cases A and B) we compare three models of the magnetic field evolution: constant field (CF), exponential decay (ED), and Hall attractor (HA); and two velocity values $v = 30$ and $100$~km~s$^{-1}$. The results of calculations for case A of propeller spin-down are presented in Fig.~\ref{ShakuraSd}. The case B is shown in Fig.~\ref{DOsd}. For illustrative purposes, we also provide a table of characteristic periods~(\ref{tab:tab}) for both cases A and B for the constant field.

% Пересчитано
\begin{table}[ht]
    \centering
    \begin{tabular}{c|c|c|c|c}
    \toprule
     \headrow \multicolumn{4}{c}{case A }&\\
     \midrule
         \headrow $B~\text{(G)},~v$~(km/s) & $10^{11},~30$ & $10^{14},~30$ & $10^{11},~100$ & $10^{14},~100$ \\ 
\midrule
$P_{\text{PA}},$ s & $3 \times 10^3$ & $3 \times 10^5$ & $5 \times 10^3$ & $5 \times 10^5$ \\ 
$P_{\text{cr}},$ s & $3 \times 10^3$ & $6 \times 10^5$ & $2 \times 10^4$ & $2 \times 10^7$ \\ 
$P_{\text{turb}},$ s & $> 5 \times 10^4$ & $6 \times 10^7$ & $> 5 \times 10^4$ & $2 \times 10^{10}$ \\ 
$t_\text{PA},$ yrs & $10^7$ & $5 \times 10^6$ & $2 \times 10^8$ & $3 \times 10^7$ \\ 
$t_{\text{cr}},$ yrs & $10^7$ & $5 \times 10^6$ & $4 \times 10^9$ & $4 \times 10^7$ \\ 
$t_{\text{turb}},$ yrs & $> 10^{10}$ & $2 \times 10^7$ & $> 10^{10}$ & $5 \times 10^9$ \\  
    %\bottomrule
    \end{tabular}

    \begin{tabular}{c|c|c|c|c}
    \toprule
     \headrow \multicolumn{4}{c}{case B }&\\
     \midrule
         \headrow $B~\text{(G)},~v$~(km/s) & $10^{11},~30$ & $10^{14},~30$ & $10^{11},~100$ & $10^{14},~100$ \\ 
\midrule
$P_\text{PA},$ s & $3 \times 10^3$ & $3 \times 10^5$ & $5 \times 10^3$ & $5 \times 10^5$ \\ 
$P_{\text{cr}},$ s & $3 \times 10^3$ & $6 \times 10^5$ & $2 \times 10^4$ & $2 \times 10^7$ \\ 
$P_\text{turb},$ s & $> 5 \times 10^4$ & $6 \times 10^7$ & $> 5 \times 10^4$ & $2 \times 10^{10}$ \\ 
$t_\text{PA},$ yrs & $6 \times 10^7$ & $2 \times 10^7$ & $2 \times 10^9$ & $2 \times 10^8$ \\ 
$t_{\text{cr}},$ yrs & $6 \times 10^7$ & $2 \times 10^7$ & $6 \times 10^9$ & $2 \times 10^8$ \\ 
$t_{\text{turb}},$ yrs & $> 10^{10}$ & $3 \times 10^7$ & $> 10^{10}$ & $5 \times 10^9$ \\ 

% старая версия ниже. Я хочу сравнить
%$P_\text{PA},$ s & $3 \times 10^3$ & $3 \times 10^5$ & $5 \times 10^3$ & $5 \times 10^5$ \\ 
%$P_{\text{cr}},$ s & $3 \times 10^3$ & $6 \times 10^5$ & $2 \times 10^4$ & $2 \times 10^7$ \\ 
%$P_{\text{turb}},$ s & $> 5 \times 10^4$ & $6 \times 10^7$ & $> 4 \times 10^4$ & $2 \times 10^{10}$ \\ 
%$t_\text{PA},$ yrs & $9 \times 10^7$ & $2 \times 10^7$ & $3 \times 10^9$ & $3 \times 10^8$ \\ 
%$t_{\text{cr}},$ yrs & $9 \times 10^7$ & $2 \times 10^7$ & $7 \times 10^9$ & $3 \times 10^8$ \\ 
%$t_{\text{turb}},$ yrs & $> 10^{10}$ & $4 \times 10^7$ & $> 10^{10}$ & $6 \times 10^9$ \\   
    \bottomrule
    \end{tabular}
    \caption{Critical moments of time and periods for two propeller deceleration models: case A (eq.~\ref{K_P1}) and case B (eq.~\ref{K_P2}) for a constant magnetic field.
    %($K_{\text{P}}=\Dot{M} \omega R_{\text{m}}^2$, \cite{1975PAZh....1...23S}) and case B ($K_{\text{P}} = \dot{M} \sqrt{GMR_{\text{m}}}$, \cite{1973ApJ...179..585D}) for a constant magnetic field. %The characteristic period of the onset of accretion $P_{\text{PA}}$, the critical period for the turbulence to become significant $P_{\text{cr}}$ and the characteristic period during the turbulent regime $P_{\text{turb}}$ with the corresponding neutron star ages $t_{\text{PA}}$, $t_{\text{cr}}$, $t_{\text{turb}}$. The field is constant.
    }
    \label{tab:tab}
\end{table}

Below we describe the properties of each stage as we get it from our calculations.

\paragraph{Ejector} In our calculations each INS starts its evolution with an initial period of $100$~s. The combination of the velocity $v$ and the initial magnetic field $B_0$ defines the first stage, since both of the transition periods $P_{\text{EP}}$ and $P_{\text{PA}}$ depend on these parameters. As it is noticed by \citet{2023arXiv230912080A}, long-period INSs with $B \lesssim 10^{13}$~G in a typical ISM cannot be at the ejector stage. For both adopted velocity values, only NSs with the magnetic field $10^{14}$~G start their evolution at the ejector stage. In all other cases, NSs start as propellers. Therefore, on every plot, the early spin-down of NSs with $B_0 = 10^{11}, 10^{12}$, and $10^{13}$~G obeys the propeller spin-down law, and the NS with the magnetic field $10^{14}$~G undergoes ejector spin-down.
% The ejector-propeller transition for $10^{14}$~G
On the other hand, all modeled NSs end their evolution as accretors.

 %The NS with $B_0=10^{14}$~G, unlike the others, starts its evolution at the ejector stage. 
 The transition to the propeller stage occurs when $P = P_{\text{EP}}$. The corresponding moment of time $t_\text{EP}$ depends on the specific combination of $v$, $B_0$, and the adopted field decay model. For instance, for models CF and ED the transition time is almost exactly the same: $t_\text{EP} = 5\times 10^6$~yrs. This is so because the characteristic decay timescale for the ED model is long~--- $\tau_{\text{Ohm}} \approx 10^9 \gg 10^6$~yrs~--- and therefore, the NSs' parameters in both cases (A and B) are similar during the ejector-propeller transition. On the contrary, for the HA model $t_\text{EP} \sim 10^6$~yrs which is comparable to the field decay timescale. So, the change to the propeller in the HA case occurs due to field decay, which leads to a decrease of $R_{\text{Sh}}$. Therefore, the condition $R_{\text{Sh}}=R_{\text{G}}$ is fulfilled earlier, since $R_{\text{G}}$ is constant.
 So, for the NS with $B_0=10^{14}$~G and the HA model (two bottom panels in both Figures) the transition to the propeller stage occurs due to the field decay rather than spin-down due to pulsar losses.

% $10^{14}$~G propeller appearance and duration 
\paragraph{Propeller} Let us consider first the propeller stage of an NS with an initial magnetic field strength of $B_0 = 10^{14}$~G. In case A, the energy loss rate of the propeller is significantly greater than that of the ejector. So, the spin-down rate changes dramatically, as indicated by the break in lines for $10^{14}$~G. In this case, within model A and for field decay models CF and ED, the duration of the propeller stage is much shorter than the duration of the ejector stage. As a result, the period increases very rapidly by approximately three orders of magnitude in a rather short time. Thus, for CF and ED and for $v = 30$~km~s$^{-1}$ the propeller stage starts at $\sim 5 \times 10^{6}$~yrs and lasts only for $\lesssim 10^{5}$~yrs, while the period $P$ changes from $140$~s to $3\times10^5$~s during this stage. On the other hand, for the HA field evolution model, the duration of this stage is comparable to that of the ejector due to magnetic field decay, which makes the deceleration less effective.

% Можно написать в такой форме, но это будет противоречить LipunovPopov1995: Для B = 10^14 Гс замедление Д&О мало. Оно сравнивается с замедлением на эжекторе только при P = 394 с (при больших P момент замедления пропеллера больше эжекторного). Это видно на втором графике, справа сверху. 10^14 Гс стартует с эжектора, а на аккретор выходит раньше всех. Для 10^13 Гс равенство достигается при 101 с. 

%After consideration of $10^{14}$~G we will consider NSs with other initial magnetic field values and compare their propeller stage to the propeller stage of NS with $B_0 = 10^{14}$~G. 
As for models with $B_0 < 10^{14}$~G, all of them start their evolution as propellers. For cases A and B, the weaker the magnetic field the slower the spin-down. 
%Therefore the spin period of NS with $B_0=10^{13}$~G is longer that that of the same age star with $B_0=10^{12}$~G or $B_0=10^{11}$~G.

Let us note the difference between the HA model and other field decay models for cases A and B. For the HA model, the characteristic decay timescale $\sim10^{6}$~yrs is comparable to  $t_{\text{PA}} \sim 2\times10^5-5\times10^6$~yrs in case A. Therefore, it affects the period evolution. On the contrary, in the case B $t_{\text{PA}} \gtrsim 2\times10^7$~yrs is much greater than the decay timescale for every $B_0$ and $v$ values considered in this work. So, the transition to the propeller stage and further evolution is almost the same as it would be with a constant field of the value $B \approx 0.05 B_0$. 
%$\gg 10^6$~yrs

Now we consider the difference of the propeller-accretor transition of NSs with $B_0 = 10^{14}$~G and NSs with smaller initial magnetic fields. For model HA of case A and all decay models of case B the ejector and propeller spin-down rate can be considered comparable. So, for larger $B_0$ the propeller-accretor transition happens earlier.  On the contrary, in models CF and ED of case A the propeller spin-down is much more effective. A NS with $B_0 = 10^{14}$~G spends part of its early evolution at the ejector stage,
%some time at the ejector stage before reaching $P_{\text{PA}}$,
while NSs with lower $B_0$ are born already at the propeller stage. This delays the propeller-accretor transition for the NS with $B_0 = 10^{14}$~G in comparison to the transition of NSs with lower magnetic fields.
 
While the value of the velocity $v$ does not affect the ejector spin-down, the ejector-propeller transition period $P_{\text{EP}} \propto v^{1/2}$, so $B_0=10^{14}$~G and either $v=30$~km~s$^{-1}$ or $100$~km~s$^{-1}$ correspond to $P_\text{EP}~=~140$~s and $260$~s, respectively. The latter makes the ejector stage longer. Thus, for field decay models CF and ED the duration of the ejector stage increases from $5 \times 10^6$~yrs to $3 \times 10^7$~yrs. At the same time, the propeller stage is due to interaction with the ISM and corresponding spin-down laws depend on $\dot{M} \propto v^{-3}$. So, the propeller spin-down rate is much less effective for high velocities. Hence, the moment of the propeller-accretor transition also depends on $v$, although this dependence is rather weak ($P_{\text{PA}} \propto v^{1/3}$). As a result, the transition to the accretor stage for all high-velocity cases occurs later than that for low-velocity ones. At the same time, lower $P_\text{PA}$ correspond to lower values of the magnetic field, as it follows from eq.~(\ref{P_PA}) if $v$ remains unchanged. Therefore, in our results, we see the accretor stage for high $B_0$ starts at a longer period $P_{\text{PA}}$.

%Now, we consider the spin period evolution for $t>t_{\text{PA}}$. During the propeller regime, the magnetosphere is surrounded by the material from the ISM. All of the matter $\dot{M}$ was added from the outer bounder of the envelope and before interaction with the magnetosphere changed its properties, so the turbulent moment it was carrying was not important. As soon as the Bondi-Hoyle-Littleton accretion begins ($t=t_{\text{PA}}, P=P_{\text{PA}}$), the material from the ISM with the turbulent moment interacts directly with the magnetosphere. So, only after $t > t_{\text{PA}}$ we are going to keep track of turbulence.

\paragraph{Bondi Accretor} Now, let us consider the spin period evolution for $t > t_{\text{PA}}$. 
%At the stage of accretion, 
If the spin period becomes long enough, we have to account for the interstellar turbulence. The turbulent regime is reached when an NS spins down to the critical period $P_{\text{cr}}$. At this moment the turbulent torque becomes comparable to the accretor spin-down torque given by eq.~(\ref{K_A}). Still, at the onset of the turbulent regime of the spin evolution, the NS is far from the equilibrium (the latter is expected since turbulence effectively spins up the star~--- see below). In Figs.~\ref{ShakuraSd} and~\ref{DOsd} the moments $t = t_{\text{cr}}$ when $P = P_{\text{cr}}$ are marked with filled circles on the curves. 

In the situation, when the period $P_{\text{cr}} < P_{\text{PA}}$ we consider the turbulence to become significant right after the accretion starts and assume $P_{\text{cr}} = P_{\text{PA}}$ and $t_{\text{cr}} = t_{\text{PA}}$. The condition $P_{\text{cr}} < P_{\text{PA}}$ is satisfied for NSs with $v = 30~\text{km~s}^{-1}$. On the other hand, the turbulent regime starts later for $v = 100~\text{km~s}^{-1}$, as can be seen from Table~\ref{tab:tab}. This is so as  $P_{\text{cr}} \propto v^{17/6}$, thus it depends much stronger on $v$ than $P_{\text{PA}} \propto v^{1/3}$.
%and becomes larger than $P_{\text{PA}}$ with increase in velocity $v$ for all values of the initial magnetic field.

% может часть из этого должна быть в модели
\paragraph{Turbulent regime} The spin period evolution after reaching the turbulent regime at $t_{\text{cr}}$ is influenced by the turbulence. How quickly the turbulent regime is established depends on the spin-down torque. In the case of settling accretion, it happens much faster, on a time scale comparable to the duration of the propeller regime under the torque defined by eq.~(\ref{K_P1}). In the case of Bondi accretion, the turbulent regime appears later. Below we present the results for this assumption.

Due to the existence of stochastic turbulent torque, the NS spin evolution can not be described only with the accretor spin-down torque, eq.~(\ref{K_A}). So, the stellar spin period obtained in our calculations after $t_\text{turb}$ can be interpreted as an upper bound value. Therefore, we actually give only a lower bound for the time ($t_{\text{turb}}$) of reaching the equilibrium. In Figs.~\ref{ShakuraSd} and~\ref{DOsd} the moments $t = t_{\text{turb}}$ when $P = P_{\text{turb}}$ are marked with triangles on the curves. 

In general, the stronger the magnetic field $B$ and the lower the velocity $v$, the earlier an NS reaches $P_{\text{turb}}$. For instance, in cases A and B for models CF and ED the NS with $B_0=10^{14}$~G reaches the equilibrium, while for lower initial magnetic field values it is not necessarily true. For example, for CF, as shown in Table~\ref{tab:tab}, the NS with the lowest magnetic field $10^{11}$~G does not reach the equilibrium period for both velocity values.

In contrast with CF and ED models, an NS within the HA model reaches the equilibrium later, despite the fact that $P_{\text{turb}}$ in these cases does not vary significantly. That is so because the spin-down rate at the accretor stage depends on the magnetic field value. After $\sim$~few$\times 10^{5}$--$\sim$~few$\times 10^{6}$~yrs the magnetic field in the HA model decreases down to $\approx 0.05 B_0$. Thus, the drop in the $B$ value results in slowing down the increase of the period. As for velocity values, $P_{\text{turb}}$ is highly dependent on $v$ ($P_{\text{turb}} \propto v^{43/9}$). So, NSs with $v = 100~\text{km~s}^{-1}$ reach the equilibrium regime later than NSs with $v = 30~\text{km~s}^{-1}$. After the equilibrium period is reached, we assume $P = P_{\text{turb}}$. For the CF model, $P_{\text{turb}}$ does not change with time, because for this model $B$ is constant. In the HA model, after $P_{\text{turb}}$ is reached, the magnetic field value is also constant because the field decays to its final value of $\approx 0.05 B_0$ in $10^{6}$~yrs which is $\ll t_{\text{turb}}$. After $t_{\text{turb}}$ the field decay continues only for the ED model. Here the noticeable decrease in $B$ occurs at a time $\sim 10^9$~yrs. This can be seen as a decrease in $P_{\text{turb}}$ in the middle panels of both Figures.

\section{Discussion}

In this paper, we aimed to study the long-term evolution of INSs with long initial spin periods $\sim 100$~s focusing on their transition to the stage of accretion. 
Inevitably, we made many simplifying assumptions. Some of them are discussed here. 

The main simplification is related to the usage of constant spatial velocities and ISM density. Previous calculations (e.g. \cite{2000ApJ...530..896P}) demonstrated that detailed calculations of the spatial evolution are important. 
Still, basic properties of the evolution of INSs with potentially larger luminosity at the accretion stage can be calculated under these assumptions, as these objects have low spatial velocities of about a few tens of km per second. So, they spend their lives close to the Galactic plane. However, the short duration of pre-accretor stages for NSs with long initial spin periods demonstrate that such objects can reach the accretor phase of evolution even if they spend just a part of their lives in a high-density environment in the Galactic disk. Our preliminary calculations indicate that NSs with standard initial periods ($P_0\lesssim1$~s) and magnetic fields ($B\sim 10^{11} - 10^{12}$~G) would not start to accrete if they spend a significant part of their lives above the Galactic plane. In the future, we plan to include details of spatial evolution in order to make a realistic population model. 

In this study, we used three very simple models of the magnetic field evolution. 
Contrary to the assumptions discussed before, here we are limited by uncertainties in the long-term field behavior. Potentially, it is possible to implement a more detailed treatment of the evolution of the magnetic field accounting for the thermal evolution which influences the Ohmic time scale, for the evolution of the field in the core, etc. Still, uncertainties are so big that we think that it is premature to apply sophisticated scenarios. 

%Here something about fallback

Now let us discuss some issues related to our treatment of spin evolution. 
At first, we used the traditional treatment of the centrifugal barrier with the critical condition $R_{\text{c}}=R_{\text{m}}$ where $R_{\text{c}}$ is calculated according to eq.~(11). 
This standard approach was recently challenged by \citet{2023MNRAS.520.4315L}. 
The author demonstrated that for dipolar magnetically aligned case the critical radius is $(2/3)^{1/3}R_{\text{c}}$. Accounting for this effect might make the propeller stage longer. However, we notice that there are many significant uncertainties in the spin-down rate at this stage. Our estimates demonstrate that in our study uncertainties in the spin-down rate are more important than the critical condition for the transition from propeller to accretor. 

% Если бы считался Д П, то аккеторов бы не было. Пример.
In our opinion, the most crucial uncertainties in calculating the time of the onset of accretion are related to the rate of spin-down at the propeller stage. We also make calculations for cases C and D. For them we obtain that the time needed for an NS to start accretion is longer than the Galactic age: $t_{\text{PA}} > t_{\text{gal}}$.

%Здесь про дозвуковой пропеллер
\citet{1981MNRAS.196..209D} suggested that accretion does not start immediately after the condition $R_{\text{c}}=R_{\text{m}}$ is reached. After a standard (supersonic) propeller there might be another stage~--- subsonic propeller. In our study, we ignore this stage. In our opinion, the heating balance in the envelope around the magnetosphere is adequately calculated in the settling accretion model. Thus, when $R_{\text{m}}<R_{\text{c}}$ in the low luminosity case realized in the case of AINSs the settling accretion model can applied. Additionally, up to our knowledge, the existence of the subsonic propeller stage is not supported by numerical calculations.

Regarding the turbulent regime of accretion, we have to note that the details of the interaction of the heated envelope with turbulent matter captured within the settling accretion are poorly known. They are far beyond the scope of our study. Thus, we leave it to future analysis.

 Now, we want to discuss some properties of AINSs. \citet{2015MNRAS.447.2817P}  proposed that AINSs can be observed as transient sources. In our model, we do not account for the transient behavior of these sources. Potentially, transient accretion can influence the spin behavior of AINSs.
 
% Касательно наблюдательных проявлений, многие НЗ не достигают равновесия, это значит, что большинство из них будут замедляться 

As we mentioned several times, already, there are numerous uncertainties in the properties and evolution of old INSs. Only observations of these sources can shed light on many issues. Unfortunately, the eROSITA survey is not completed and all observations with this instrument are stopped. In the near future, important results can be obtained thanks to numerous discoveries of INSs via microlensing with the Roman Telescope and other instruments, including Gaia (see e.g., \cite{2020ApJ...889...31L} and references therein). Already many candidates for isolated compact objects are known (see e.g., \cite{2021arXiv210713697M} and references therein). 
Deep X-ray and infrared observations of these sources by new space observatories can provide important upper limits or even detections. 

Non-detection of AINSs even with the next generation of instruments can be explained by two reasons. Either INSs very seldom reach the stage of accretion, or the luminosity at this stage is lower than expected. Regime of low accretion with $\dot M \lesssim 10^{13}$~g~s$^{-1}$ potentially can be studied in binaries. In particular, in systems similar to those identified recently as binaries with invisible components with $M\sim 1.3-2.3 \, M_\odot$ (\cite{2022MNRAS.517.4005M, 2022arXiv220700680A}). If this question is clarified and if luminosities are not sufficiently low to explain the non-detection then the problem might be in the magneto-rotational evolution of INSs. If sources are not found then there can be several ways to explain it. For example, in the case of AINSs it can be due to properties of the propeller stage or/and due to magnetic field evolution. However, our results suggest that INSs with long initial spin periods can become accretors for various extreme variants of field behavior. Thus, we suggest that the most probable evolutionary explanation for non-detection might be related to a prolonged propeller stage. Still, let us hope that AINSs will be identified shortly. This will give us lots of important information about the properties of NSs.

% fallback origin of long-period pulsars, toy model does not work
% questionable velocity, velocity estimate for PSR 76 s z/v=t
% our propeller has a high energy loss rate -> early accretion
% accretion at the propeller stage (to Accretor paragraph)
% accretion due turbulence j_t = j_K is not possible
% PSR with 20 min period 
% slow deceleration while R_{\text{m}} = R_{\text{l}} and other types of the propeller stages

% Fallback - low-velocity - high luminosity 

%20-minute PSRs - the can be Ejectors only for very large field and velocities!

%NSs found with microlensing searches can used as possible targets for X-ray and IR observations

%Better understanding of accretion physics and NS evolution might allow us to make better constraints even from negative results of X/IR searches.
%See Lam et al. (2020) ApJ, 889, 31,

\section{Conclusion}

 In this paper, we study the long-term behavior of isolated NSs with long ($\sim 100$~s) initial spin periods for a range of initial magnetic fields, for different models of the field evolution, two values of the spatial velocity relative to the ISM, and constant external conditions. Our main goal is to calculate if INSs with such parameters can reach the stage of accretion of the interstellar gas within the Galactic age. 

 We find that the result strongly depends on the description of spin-down during the propeller stage. For realistic models with a rapid spin evolution at this stage (\cite{1973ApJ...179..585D, 1975SvAL....1..223S}), all INSs with reasonably low velocities (which potentially make it possible to have a relatively bright accretor) reach the stage of accretion in less than a few billion years. 

 Thus, we predict that if objects similar to PSR J0901-4046 (and, probably, GLEAM-X J1627-52 and GPM J1839-10) form a significant population of INSs in the Galaxy, then the expected number of isolated accretors is sufficiently increased in comparison to the predictions made in earlier studies. 

\begin{acknowledgement}
We thank Prof. N.I. Shakura for discussions and the anonymous referee for useful comments and suggestions.
\end{acknowledgement}

\paragraph{Funding Statement}

The work of M.A. and A.B. was supported by the RSF grant 21-12-00141.
S.P. acknowledges support from the Simons Foundation.

%This research was supported by grants from the <funder-name> <doi> (<award ID>); <funder-name> <doi> (<award ID>).

\paragraph{Competing Interests}
 `None'.

\paragraph{Data Availability Statement}

The code used for modeling is accessible on request. Please, contact the first author.

%A statement about how to access data, code and other materials allowing users to understand, verify and replicate findings --- e.g. Replication data and code can be found in Harvard Dataverse: \verb+\url{https://doi.org/link}+.

%\endnote in some journals will behave like \footnote; and \printendnotes will not output anything. 
\printendnotes

\printbibliography

%\appendix

%\section{Example Appendix Section}

%sample  text 

\end{document}